\documentclass[11pt]{article}
\usepackage{blois2002,epsfig}
\def\lline#1{\hbox to\textwidth{#1}}
\newbox\hdbox%
\newcount\hdrows%
\newcount\multispancount%
\newcount\ncase%
\newcount\ncols%
\newcount\nrows%
\newcount\nspan%
\newcount\ntemp%
\newcount\mscount%
\newdimen\hdsize%
\newdimen\newhdsize%
\newdimen\parasize%
\newdimen\spreadwidth%
\newdimen\thicksize%
\newdimen\thinsize%
\newdimen\tablewidth%
\newif\ifcentertables%
\newif\ifendsize%
\newif\iffirstrow%
\newif\iftableinfo%
\newtoks\dbt%
\newtoks\hdtks%
\newtoks\savetks%
\newtoks\tableLETtokens%
\newtoks\tabletokens%
\newtoks\widthspec%
\tableinfotrue%
\catcode`\@=11%
\chardef\@ne=1
\chardef\tw@=2%
\countdef\m@ne=22 \m@ne=-1%
\def\tstrut{\vrule height3.1ex depth1.2ex width0pt}%
\def\and{\char`\&}%
\def\tablerule{\noalign{\hrule height\thinsize depth0pt}}%
\def\thickrule{\noalign{\hrule height\thicksize depth0pt}}%
\def\hrulefill{\leaders\hrule\hfill}%
\def\ctr#1{\hfil\ #1\hfil}%
\tablewidth=-\maxdimen%
\spreadwidth=-\maxdimen%
\def\tabskipglue{0pt plus 1fil minus 1fil}%
\centertablestrue%
\gdef\ARGS{########}%
\gdef\headerARGS{####}%
\def\@mpersand{&}%
{\catcode`\|=13%
\gdef\letbarzero{\let|0}%
\gdef\letbartab{\def|{&&}}%
\gdef\letvbbar{\let\vb|}%
}%
{\catcode`\&=4%
\def\ampskip{&\omit\hfil&}%
\catcode`\&=13%
\let&0%
\xdef\letampskip{\def&{\ampskip}}%
\gdef\letnovbamp{\let\novb&\let\tab&}
}%
\def\begintable{%
   \begingroup%
   \catcode`\|=13\letbartab\letvbbar%
   \catcode`\&=13\letampskip\letnovbamp%
   \def\multispan##1{%
      \omit \mscount##1%
      \multiply\mscount\tw@\advance\mscount\m@ne%
      \loop\ifnum\mscount>\@ne \sp@n\repeat%
   }%
\def\sp@n{\span\omit\advance\mscount\m@ne}

   \def\|{%
      &\omit\widevline&%
   }%
   \ruledtable%
}%
\long\def\ruledtable#1\endtable{%
   \offinterlineskip%
   \tabskip 0pt%
   \def\widevline{\vrule width\thicksize}%
   \def\endrow{\@mpersand\omit\hfil\crnorm\@mpersand}%
   \def\crthick{\@mpersand\crnorm\thickrule\@mpersand}%
   \def\crthickneg##1{\@mpersand\crnorm\thickrule
          \noalign{{\skip0=##1\vskip-\skip0}}\@mpersand}%
   \def\crnorule{\@mpersand\crnorm\@mpersand}%
   \def\crnoruleneg##1{\@mpersand\crnorm
          \noalign{{\skip0=##1\vskip-\skip0}}\@mpersand}%
   \let\nr=\crnorule%
   \def\endtable{\@mpersand\crnorm\thickrule}%
   \let\crnorm=\cr%
   \edef\cr{\@mpersand\crnorm\tablerule\@mpersand}%
   \def\crneg##1{\@mpersand\crnorm\tablerule
          \noalign{{\skip0=##1\vskip-\skip0}}\@mpersand}%
   \let\ctneg=\crthickneg
   \let\nrneg=\crnoruleneg
   \the\tableLETtokens%
   \tabletokens={&#1}%
   \countROWS\tabletokens\into\nrows%
   \countCOLS\tabletokens\into\ncols%
   \advance\ncols by -1%
   \divide\ncols by 2%
   \advance\nrows by 1%
   \iftableinfo %
      \immediate\write16{[Nrows=\the\nrows, Ncols=\the\ncols]}%
   \fi%
   \ifcentertables
      \ifhmode \par\fi%
      \lline{%
      \hss%
   \else %
      \hbox{%
   \fi
      \vbox{%
         \makePREAMBLE{\the\ncols}%
         \edef\next{\preamble}%
         \let\preamble=\next%
         \makeTABLE{\preamble}{\tabletokens}%
      }%
      \ifcentertables \hss}\else }\fi%
   \endgroup%
   \tablewidth=-\maxdimen%
   \spreadwidth=-\maxdimen%
}%
\def\makeTABLE#1#2{%
   {%
   \let\ifmath0%
   \let\header0%
   \let\multispan0%
   \ncase=0%
   \ifdim\tablewidth>-\maxdimen \ncase=1\fi%
   \ifdim\spreadwidth>-\maxdimen \ncase=2\fi%
   \relax%
   \ifcase\ncase %
      \widthspec={}%
   \or %
      \widthspec=\expandafter{\expandafter t\expandafter o%
                 \the\tablewidth}%
   \else %
      \widthspec=\expandafter{\expandafter s\expandafter p\expandafter r%
                 \expandafter e\expandafter a\expandafter d%
                 \the\spreadwidth}%
   \fi %
   \xdef\next{%
      \halign\the\widthspec{%
      #1%
      \noalign{\hrule height\thicksize depth0pt}
      \the#2\endtable%
      }%
   }%
   }%
   \next%
}%
\def\makePREAMBLE#1{%
   \ncols=#1%
   \begingroup%
   \let\ARGS=0%
   \edef\xtp{\widevline\ARGS\tabskip\tabskipglue%
   &\ctr{\ARGS}\tstrut}%
   \advance\ncols by -1%
   \loop%
      \ifnum\ncols>0 %
      \advance\ncols by -1%
      \edef\xtp{\xtp&\vrule width\thinsize\ARGS&\ctr{\ARGS}}%
   \repeat
   \xdef\preamble{\xtp&\widevline\ARGS\tabskip0pt%
   \crnorm}%
   \endgroup%
}%
\def\countROWS#1\into#2{%
   \let\countREGISTER=#2%
   \countREGISTER=0%
   \expandafter\ROWcount\the#1\endcount%
}%
\def\ROWcount{%
   \afterassignment\subROWcount\let\next= %
}%
\def\subROWcount{%
   \ifx\next\endcount %
      \let\next=\relax%
   \else%
      \ncase=0%
      \ifx\next\cr %
         \global\advance\countREGISTER by 1%
         \ncase=0%
      \fi%
      \ifx\next\endrow %
         \global\advance\countREGISTER by 1%
         \ncase=0%
      \fi%
      \ifx\next\crthick %
         \global\advance\countREGISTER by 1%
         \ncase=0%
      \fi%
      \ifx\next\crnorule %
         \global\advance\countREGISTER by 1%
         \ncase=0%
      \fi%
      \ifx\next\crthickneg %
         \global\advance\countREGISTER by 1%
         \ncase=0%
      \fi%
      \ifx\next\crnoruleneg %
         \global\advance\countREGISTER by 1%
         \ncase=0%
      \fi%
      \ifx\next\crneg %
         \global\advance\countREGISTER by 1%
         \ncase=0%
      \fi%
      \ifx\next\header %
         \ncase=1%
      \fi%
      \relax%
      \ifcase\ncase %
         \let\next\ROWcount%
      \or %
         \let\next\argROWskip%
      \else %
      \fi%
   \fi%
   \next%
}
\def\counthdROWS#1\into#2{%
\dvr{10}%
   \let\countREGISTER=#2%
   \countREGISTER=0%
\dvr{11}%
\dvr{13}%
   \expandafter\hdROWcount\the#1\endcount%
\dvr{12}%
}%
\def\hdROWcount{%
   \afterassignment\subhdROWcount\let\next= %
}%
\def\subhdROWcount{%
   \ifx\next\endcount %
      \let\next=\relax%
   \else%
      \ncase=0%
      \ifx\next\cr %
         \global\advance\countREGISTER by 1%
         \ncase=0%
      \fi%
      \ifx\next\endrow %
         \global\advance\countREGISTER by 1%
         \ncase=0%
      \fi%
      \ifx\next\crthick %
         \global\advance\countREGISTER by 1%
         \ncase=0%
      \fi%
      \ifx\next\crnorule %
         \global\advance\countREGISTER by 1%
         \ncase=0%
      \fi%
      \ifx\next\header %
         \ncase=1%
      \fi%
\relax%
      \ifcase\ncase %
         \let\next\hdROWcount%
      \or%
         \let\next\arghdROWskip%
      \else %
      \fi%
   \fi%
   \next%
}%
{\catcode`\|=13\letbartab
\gdef\countCOLS#1\into#2{%
   \let\countREGISTER=#2%
   \global\countREGISTER=0%
   \global\multispancount=0%
   \global\firstrowtrue
   \expandafter\COLcount\the#1\endcount%
   \global\advance\countREGISTER by 3%
   \global\advance\countREGISTER by -\multispancount
}%
\gdef\COLcount{%
   \afterassignment\subCOLcount\let\next= %
}%
{\catcode`\&=13%
\gdef\subCOLcount{%
   \ifx\next\endcount %
      \let\next=\relax%
   \else%
      \ncase=0%
      \iffirstrow
         \ifx\next& %
            \global\advance\countREGISTER by 2%
            \ncase=0%
         \fi%
         \ifx\next\span %
            \global\advance\countREGISTER by 1%
            \ncase=0%
         \fi%
         \ifx\next| %
            \global\advance\countREGISTER by 2%
            \ncase=0%
         \fi
         \ifx\next\|
            \global\advance\countREGISTER by 2%
            \ncase=0%
         \fi
         \ifx\next\multispan
            \ncase=1%
            \global\advance\multispancount by 1%
         \fi
         \ifx\next\header
            \ncase=2%
         \fi
         \ifx\next\cr       \global\firstrowfalse \fi
         \ifx\next\endrow   \global\firstrowfalse \fi
         \ifx\next\crthick  \global\firstrowfalse \fi
         \ifx\next\crnorule \global\firstrowfalse \fi
         \ifx\next\crnoruleneg \global\firstrowfalse \fi
         \ifx\next\crthickneg  \global\firstrowfalse \fi
         \ifx\next\crneg       \global\firstrowfalse \fi
      \fi%
\relax%
      \ifcase\ncase %
         \let\next\COLcount%
      \or %
         \let\next\spancount%
      \or %
         \let\next\argCOLskip%
      \else %
      \fi %
   \fi%
   \next%
}%
\gdef\argROWskip#1{%
   \let\next\ROWcount \next%
}%
\gdef\arghdROWskip#1{%
   \let\next\ROWcount \next%
}%
\gdef\argCOLskip#1{%
   \let\next\COLcount \next%
}%
}%
}%
\def\spancount#1{%
   \nspan=#1\multiply\nspan by 2\advance\nspan by -1%
   \global\advance \countREGISTER by \nspan
   \let\next\COLcount \next}%
\def\dvr#1{\relax}%
\def\header#1{%
\dvr{1}{\let\cr=\@mpersand%
\hdtks={#1}%
\counthdROWS\hdtks\into\hdrows%
\advance\hdrows by 1%
\ifnum\hdrows=0 \hdrows=1 \fi%
\dvr{5}\makehdPREAMBLE{\the\hdrows}%
\dvr{6}\getHDdimen{#1}%
{\parindent=0pt\hsize=\hdsize{\let\ifmath0%
\xdef\next{\valign{\headerpreamble #1\crnorm}}}\dvr{7}\next\dvr{8}%
}%
}\dvr{2}}%
\def\makehdPREAMBLE#1{%
\dvr{3}%
\hdrows=#1%
{
\let\headerARGS=0%
\let\cr=\crnorm%
\edef\xtp{\vfil\hfil\hbox{\headerARGS}\hfil\vfil}%
\advance\hdrows by -1%
\loop%
\ifnum\hdrows>0%
\advance\hdrows by -1%
\edef\xtp{\xtp&\vfil\hfil\hbox{\headerARGS}\hfil\vfil}%
\repeat%
\xdef\headerpreamble{\xtp\crcr}%
}
\dvr{4}}%
\def\getHDdimen#1{%
\hdsize=0pt%
\getsize#1\cr\end\cr%
}%
\def\getsize#1\cr{%
\endsizefalse\savetks={#1}%
\expandafter\lookend\the\savetks\cr%
\relax \ifendsize \let\next\relax \else%
\setbox\hdbox=\hbox{#1}\newhdsize=1.0\wd\hdbox%
\ifdim\newhdsize>\hdsize \hdsize=\newhdsize \fi%
\let\next\getsize \fi%
\next%
}%
\def\lookend{\afterassignment\sublookend\let\looknext= }%
\def\sublookend{\relax%
\ifx\looknext\cr %
\let\looknext\relax \else %
   \relax
   \ifx\looknext\end \global\endsizetrue \fi%
   \let\looknext=\lookend%
    \fi \looknext%
}%
\def\tablelet#1{%
   \tableLETtokens=\expandafter{\the\tableLETtokens #1}%
}%

\newcount\figurecount     \figurecount=0
\def\FIG#1{ \global\advance\figurecount by 1 \xdef#1{\the\figurecount}}

\let\cl=\centerline
\def\allcap#1;#2;{{\renewcommand{\baselinestretch}{.8}\captionfont
\newdimen\fcwidth \fcwidth=\textwidth \advance\fcwidth by -2cm
\setbox0=\hbox{{\bf Fig. #1.} #2}  
  \ifdim \wd0>\fcwidth  
       \vbox{\noindent
          \parshape=1 1truecm \fcwidth {\bf Fig. #1.} #2}
    \else
       \cl{{\bf Fig. #1.} #2}
    \fi} }

\def\figbox#1;#2;{\parbox{#2}{\epsfig{file=#1.eps,width=#2}}}
\def\allfig#1;#2;#3;#4;{\vskip4mm\vbox{ 
  \cl{\parbox{#2 cm}{\epsfig{file=\figdir #1.eps,width=#2 cm}}} 
   \vskip2mm
\allcap#3; #4;
\vskip1mm\noindent\kern-0pt}}

\def\ie{{\it\kern-2pt i.\kern-.5pt e.\kern-2pt}}  
\def\BR{\hbox{BR}}  
\def\up#1{$^{#1}$}  
\def\ifm#1{\relax\ifmmode#1\else$#1$\fi}
\def\Bbar{\ifm{\rlap{\kern.22em\raise1.9ex\hbox to.58em{\hrulefill}} B}}

\def\to{\ifm{\rightarrow}} \def\sig{\ifm{\sigma}}   
\def\K{\ifm{K}}  
\def\ff{$\phi$--factory}  \def\DAF{DA\char8NE}  \def\f{\ifm{\phi}} 
 \def\pic{\ifm{\pi^+\pi^-}} \def\pio{\ifm{\pi^0\pi^0}} 
  
\def\po{\ifm{\pi^0}}
\def\ks{\ifm{K_S}} \def\kl{\ifm{K_L}} 
  
\def\eps{\ifm{\epsilon}} \def\epm{\ifm{e^+e^-}}
\def\rep{\ifm{\Re(\eps'/\eps)}}    
\def\Kb{\ifm{\rlap{\kern.3em\raise1.9ex\hbox to.6em{\hrulefill}} K}}
\def\kpm{\ifm{K^\pm}}  
\def\C{\ifm{C}}  \def\P{\ifm{P}}  \def\T{\ifm{T}}
\def\noc{\relax\hglue0pt{\rlap{$C$}\raise.15ex\hbox{$\kern
.18em\backslash$}}}
\def\nop{\relax\hglue0pt{\rlap{$P$}\raise.15ex\hbox{$\kern
.18em\backslash$}}}
\def\noT{\relax\hglue0pt{\rlap{$T$}\raise.15ex\hbox{$\kern
.18em\backslash$}}}
\def\nocp{\noc\nop} 
\def\ko{\ifm{K^0}}  \def\kob{\ifm{\Kb\vphantom{K}^0}}
\def\gam{\ifm{\gamma}}  
 \def\ab{\ifm{\sim}}  \def\x{\ifm{\times}}
\def\sta#1{\ifm{|\,#1\,\rangle}} 
\def\L{\ifm{{\cal L}}}  
\def\pt#1,#2,{\ifm{#1\x10^{#2}}}
\def\kon{\ifm{K_1}} \def\ktw{\ifm{K_2}} 
  \def\dif{\hbox{d}}   \def\Gam{\ifm{\Gamma}}
\def\bye{\end{document}}
\def\bbb{\begin{document}}
\let\cl=\centerline

\def\qq{$q\bar q$} \def\qqq{$qqq$}  \def\BR{\hbox{BR}}
\def\pppco{\ifm{\pi^+\pi^-\pi^-}}  \def\pppo{\ifm{\pi^0\pi^0\pi^0}}
\def\figbox#1;#2;{\parbox{#2cm}{\epsfig{file=\figdir #1.eps,width=#2cm}}}
\def\cfigbox#1;#2;{\cl{\figbox #1;#2;}}
\def\lam{\ifm{\lambda}}

\makeatletter
\newskip\z@skip \z@skip=0pt plus0pt minus0pt
\def\m@th{\mathsurround=\z@}
\def\ialign{\everycr{}\tabskip\z@skip\halign} 
\def\eqalign#1{\null\,\vcenter{\openup\jot\m@th
\makeatother  \ialign{\strut\hfil$\displaystyle{##}$&$\displaystyle{{}##}$\hfil
      \crcr#1\crcr}}\,}
\def\figdir{}

\begin{document}
\vspace*{4cm}
\title{CP AND CPT STUDIES WITH KAONS}

\author{JULIET LEE-FRANZINI\hglue1mm\footnote{\it{e-mail:Juliet.Leefranzini@LNF.INFN.IT}}}

\address{Laboratori Nazionali di Frascati, Via E. Fermi 40, Frascati, I-00044 Italy}

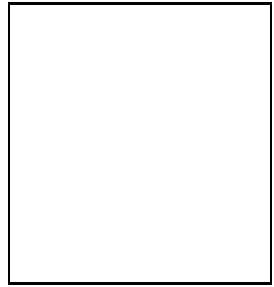
\maketitle\abstracts{Three European experiments are engaged at present in the study of \C\P\ violation in the kaon system. In particular, NA48/1 and KLOE are concerned with the \K-short decaying into three pions, while NA48/2 and KLOE study the charged kaon decaying into three pions' Dalitz plot asymmetries. I will discuss the physics involved as well as summarize the anticipated sensitivities which can be reached by the experiments.
}

\def\captionfont{\small}\FIG\Dsdq
\FIG\Ktria
\FIG\Kconst
\section{Introduction}

I will precede my discussion of what is expected, from on-going and future experiments on \C\P\ violation and \C\P\T\ tests, with a few historical reminders, to place the present measurements in perspective. The kaons were discovered around 1947, as a surprise, and since then, they have been a cornucopia for particle physics. We owe to them the ideas of ``flavor'',``quark'', and ``flavor mixing''.  Parity violation first showed up in the ``$\tau - \theta$'' puzzle. From semi-leptonic decays we learnt of the $\Delta S=\Delta Q$ rule and obtained the first value of $\sin\theta_{\rm C}$. The kaons' two pion decays taught us the dominance of $\Delta I=1/2$ transitions, whose explanation is still being labored over by lattice and chiral perturbation physicists. Finally, of course, \C\P\ violation, \noc\nop\ in the following. Indirect \C\P\ violation had been discovered long ago and direct \C\P\ violation has only just been definitively established, in the neutral kaon system.

Just for nostalgia, I list some ``recent'' historical dates: (1) 1964 \ \nocp\ in $|\Delta S|=2$ seen. (2) 1967-1973 \ ${\cal A}^L_\ell\ab\sqrt2\Re\eps$. (3) 1970-1980 \ $\left|V_{us}\right|$=0.220 $\pm$1.1\% experimentally. (4) 2001 \ Direct \nocp\ in $|\Delta S|=1$ established. (5) 2001 \ \nocp\ in $|\Delta B|=2$ seen. (6) 1960-Today \ $M(\kl)-M(\ks)$, \eps, \rep\  etc, accuracies are being improved. (7) 1990's \ $|M(\ko)-M(\kob)|<\pt4.5,-10,$ eV.

\section{Other \nocp\ Kaon Physics}

In the near future, three experiments in Europe, NA48/1~\cite{NA48a},  NA48/2~\cite{NA48b}, and KLOE, are set to do the first five measurements listed below. The sixth needs such high intensity kaon beams that given the fiscal stringency at CERN (so that its proposed high-intensity proton driver~\cite{hpd} is not likely to be built), probably can only be realized in the U. S., if KOPIO at BNL is funded.
(1) \ks\to\pio\po, \BR\ab\pt2,-9,: NA48/1, KLOE 2004. (2) Odd pion slopes from \K\up+-\K\up-: NA48/2, KLOE 2004. (3) \kpm\to$\pi^\pm$\po\gam, NA48/2, KLOE 2004. (4) $\Gamma(K^+\to\pi^+\pi^+\pi^-)-\Gamma(K^-\to\pi^-\pi^-\pi^+)$: KLOE 2004. (5) \ks\to\po$e^+e^-$: NA48/1, KLOE 2004. (6) \kl\to\po$\nu\bar\nu$: \BR\ab\pt3,-11,, KOPIO 20??. The above list is in order of experimental difficulty and descending branching ratios, \BR's. The expected signals are quite predictable theoretically for items 1 and item 6, while the situation is different for items 2, 3 and 4. Item 5 is an ancillary measurement, useful for isolating the direct \C\P\ violating component in \kl\to\po$e^+e^-$. 

\subsection{\ks\to\pio\po}

Since there has been some confusion in what this \BR\ should be lately,
I will write down a couple of lines on its derivation.\\
Let: \sta\ks=\sta\kon+\eps\sta\ktw; \kl=\sta\ktw+\eps\sta\kon; $\eta_i=\langle i\sta\kl/\langle i\sta\ks$; $\eps=(2\eta_{+-}+\eta_{00})/3.$ Then:
$$\eta_{000}={\langle3\po\sta\kl\over\langle3\po\sta\ks}
=\eps+\eps'_{000}; \quad|\eps'_{000}/\eps|\ll1$$
$${\BR(\ks\to3\po)=|\eta_{000}|^2\x\BR(\kl\to3\po)\x
{\Gamma_L\over\Gamma_S}=|\eps|^2\x\BR(\kl\to3\po)\x
{\tau_S\over\tau_L}}$$
Putting in the presently known values of \eps, the \kl\to 3\po \ branching ratio and the lifetimes of \ks\ and \kl, the branching ratio for the \noc\nop\ decay of \ks\ to 3 neutral pions is \BR(\ks\to3\po)=\pt1.9,-9, with an uncertainty of \ab1.3\%.

The best experimental information to date is from a test run for NA48/1. They have no signal with preliminary values of the uncertainties of:  $$\delta\Re\,\eta_{000}\ab\pt2.2,-2,,\ \delta\Im\,\eta_{000}\ab\pt2.8,-2,,$$ from which it follows that $\delta\BR(\ks\to3\po)\ab\pt4.6,-7,$ or $\BR<\pt0.8,-6,$ at 90\% cl.

\subsection{ $\K^\pm\to3\pi$}
Both the partial decay rate and slope parameter, $g$, asymmetries in three pion charged kaon decays yield information on \C\P\ violation. In fact, they offer new means to detect ``direct \C\P\ violation". Unfortunately, the expected effects are woefully small. For instance, observation of $\Gamma(K^+\to\pi^+\pi^+\pi^-)\ne\Gamma(K^-\to\pi^-\pi^+\pi^-)$ implies direct \C\P-violation.

There are four \nocp\ asymmetries:
$$\eqalign{{\cal A}_\Gamma&={\Delta\Gamma\over2\Gamma}=
{\Gamma(\K^+\to3\pi)-\Gamma(\K^-\to3\pi)\over\Gamma(\K^+\to3\pi)+ \Gamma(\K^-\to3\pi)}\cr
{\cal A}_g&={\Delta g\over2g}=
{g(\K^+\to3\pi)-g(\K^-\to3\pi)\over g(\K^+\to3\pi)+g(\K^-\to3\pi)}\cr}$$
for both $\tau^\pm$, \ie\ $\pi^\pm\pi^+\pi^-$\ and $\tau'^\pm$ or $\pi^\pm\pio$.

The asymmetry arises from the interference of two $\Delta I=1/2$ amplitudes $a$ and $b$. There is no $\Delta I=3/2$ suppression, therefore 
$|\Im a/\Re a|$\ab$|\Im b/\Re b|$\ab10\up{-4}. However, to lowest order in chiral perturbation $\arg a=\arg b$. Asymmetries in SM are therefore very small. For example:
$${\cal A}_g=\left({\Im b\over\Re b}-{\Im a\over\Re a}\right)\sin(\alpha_0-\beta_0)={\cal O}(10^{-6})$$
where $\alpha_0$ and $\beta_0$ are small rescattering phases, and $\sin(\alpha_0-\beta_0)\ab0.1$.
From Maiani and Paver~\cite{Miapav}:
$$\eqalign{{\cal A}_g,\tau&=\pt(-2.3\pm0.6),-6,\cr
{\cal A}_g,\tau'&=\pt(1.3\pm0.4),-6,\cr
{\cal A}_\Gamma,\tau&=\pt(-6\pm2),-8,\cr
{\cal A}_\Gamma,\tau'&=\pt(2.4\pm0.8),-8,\cr}$$

But where things are small, big surprises might hide. D'Ambrosio, Isidori, and Martinelli~\cite{Damisimar} state that large \nocp\ effects, $A_g$ of ${\cal O}$(10\up{-4}), could be triggered by a misalignment of quark and squark mass matrices through the chromomagnetic operator - CMO: {\it possible only if several conditions conspire in the same direction}. Fine tuning becomes then necessary for explaining the experimental value of \rep.
\
                                                                  \section{\ks\to$\pi\ell\nu$}

From this reaction one can learn about $\Delta S=\Delta Q$ and \T\C\P\ violation by measuring (a) the rate $\Gamma(\ks\to\pi\ell\nu)$ and (b) the leptonic asymmetry ${\cal A}^S_\ell$.
\subsection{$\Delta S=\Delta Q$}
There is no $\Delta S=-\Delta Q$ in the SM. We only have $s\to W^-u$, $\bar s\to W^+ \bar u$.
The diagram illustrates a fake $\Delta S=-\Delta Q$ process, if we were just to look at the final lepton sign. At the $su$ vertex $\Delta S=\Delta Q$.
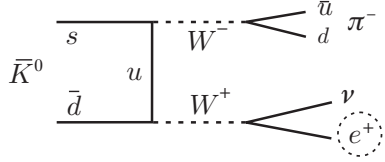
\allfig dsdqnotbw;5;\Dsdq;\kob\ decay to a ``wrong'' sign lepton.;
\noindent
From fig. \Dsdq, the amplitude ratio $x_{\ell3}=A(\bar K\to \ell^+\pi^-\nu)/A(\bar K\to \ell^-\pi^+\bar\nu)$, is ${\mathcal O} (Gm^2)$\ab10\up{-6}, for reasonable values of $m$\ab$m_K-m_\pi$.
Experimentally: $x_{\ell3}<$10\up{-2} @ 90\% CL, thus its measurement must be improved

\subsection{$TCP$, $\Delta S=\Delta Q$ and leptonic asymmetry}

We define the ``semileptonic'' asymmetry in $K_{S,L-\ell3}$ decays as
${\cal A}^{S,L}_\ell=(N(e^+)-N(e^-)/(N(e^+)+N(e^-))$.
From \T\C\P\ and $\Delta S=\Delta Q$ it follows that ${\cal A}^S_\ell={\cal A}^L_\ell$. It is not possible however to disentangle the effects of $TCP$ and $\Delta S=\Delta Q$ within the \ks-\kl\ system. It is necessary to combine results from \ko\ (or \kob) states, tagged by strong interactions. One needs {\it eg}, \epm\to\f\to$K^+K^-$. One \kpm\ tags the other. Charge exchange, in any material, gives \ko (or \kob).

\noindent
\T\C\P\ can be violated in mass-matrix and/or decay amplitudes. There are 5 complex parameters for \K\to$\pi\ell\nu$: $\delta$, the $\Delta S=\Delta Q$ amplitudes $a, b$ and the $\Delta S=-\Delta Q$ amplitudes $c,d$ satisfying:
$$\eqalign{ 2\delta& =\eps_S-\eps_L\cr
            \Re(a,c)&\qquad \hbox{\T\C\P--even}\cr
            \Im(a,c)&\qquad \hbox{\T\C\P--even}\cr
            \Re(b,d)&\qquad \hbox{\T\C\P--odd}\cr
            \Im(b,d)&\qquad \hbox{\T\C\P--odd}\cr
}$$
If $c=d=0$, then ${\cal A}^S_\ell-{\cal A}^L_\ell=4\Re\delta$.
A limit from the above improves the determination of $\left(M(\ko)-M(\kob)\right)/M$. Do note that one needs many \x10\up{10} \K's, therefore tens of fb\up{-1} at a \ff, not easy to come by.

\section{\C\P, Unitarity and Triangles}
The proverbial price of \nocp: $J=A^2\lam^6\eta$=\pt(2.7\pm1.1),-5,, is poorly known. $J$ is also (2\x)area of all unitary triangles. We must check the closing of all triangles and compare their areas. There are still many measurements needed for $B$'s, and two more for $K$'s, ignoring $D$'s for the moment.
We use in the following Wolfenstein's notation. The mixing martix parameters are
\lam=0.22 to \ab1\%, $A$\ab0.84$\pm$0.09, $|\rho-i\eta|\ab0.3\pm0.15$.

\def\figboxc#1;#2;{\vglue3mm\noindent\cl{ \epsfig{file=\figdir #1.eps,width=#2truecm}}}
\subsection{The $K$ Triangle}
\figboxc trianglek;10;
\allcap \Ktria; The unitary triangle for kaons.;
The error on $J_{12}$ is given by
$${\delta J_{12}\over J_{12}}=6\%\oplus{\delta\eta\over\eta}$$
The decay rate for $\kl\to\po\nu\bar\nu$, measures directly $\eta$:
$\Gamma(\kl\to\po\nu\bar\nu)\propto\eta^2$ and 
$J_{12}=\lambda(1-\lambda^2/2)\Im(V_{td}V_{ts}^* )\cong
5.6[\BR(\kl\to\po\nu\bar\nu)]^{1/2}$.
Thus 100 events determine $\delta\eta/\eta$ to 5\% and $J_{12}$ to \ab8\%.
More measurements, indicated in fig. \Kconst, do over-constrain the $\rho-\eta$
determination.\vglue2mm
\figboxc kconstrbw;6.3;
\allcap \Kconst; Constraints for kaons.;\vglue.1mm\noindent
So far there are only two $K^+\to\pi^+\nu\bar\nu$ events, in agreement with BR\ab\pt1.5,-10, and only hopes of observing \kl\to\po$\nu\bar\nu$.
\subsection{Unitarity}
The most stringent proof of unitarity of the mixing matrix so far is the extended GIM cancellation in $M(\kl)-M(\ks)$ and $\ko\to\mu\mu$. Additional testing should not be limited to the closing of a triangle, but of all triangles, and verifying the equality of the areas of all triangles.
\section{\DAF\ and KLOE}
The descriptions of NA48/1 and NA48/2 were given in glorious detail in the talks by Richard Batley and Gianmaria Collazuol respectively and appear in these proceedings. However, \DAF\ the particle accelerator and KLOE, its main detector, are part of my talk, hence their goals are included here.
The relevant \DAF\ parameters are given in table 1.
\newpage
\def\tstrut{\vrule height 2.4ex depth .7ex}
\cl{Table 1. \DAF\ parameters.}\vglue2mm
\begintable
 Parameter | Design | 2001 | 2004 \cr
 Bunches | 120 | 45 | 120 \cr
 Current (A) | 5 | 1.2 | \cr
 \L\ ($\mu$b\up{-1} s\up{-1}) | \pt5,32, | \pt5,31, | \pt5,33, \cr
 Beam $\tau$ (m) | $\gg$100 | $<$20 | \cr
 $\int\L\dif$t pb\up{-1} | 5000 | \ab190 | 50,000 \endtable

\subsection{The uniqueness of a \ff}
In the reaction {\epm\to\f\to}neutral kaons, the initial state is
$$\sta{i}={\sta{\ko,{\bf p}}\sta{\kob,-{\bf p}}-
\sta{\kob,{\bf p}}\sta{\ko,-{\bf p}}\over\sqrt2}$$
Defining 
$\sta{\ks}\!\equiv\!p'\sta{\ko}+q'\sta{\kob}$ and 
$\sta{\kl}\!\equiv\!p\sta{\ko}-q\sta{\kob}$ with
$|p'|^2+|q'|^2=1$ and $|p|^2+|q|^2=1$
we can also write
$$\sta{i}={\sta{\ks,{\bf p}}\sta{\kl,-{\bf p}}-
\sta{\kl,{\bf p}}\sta{\ks,-{\bf p}}\over\sqrt{2(qp'+q'p)}}$$
Therefore at \DAF\ one has pure \kl, \ks, \ko, \kob\ beams. 
From unitarity and $\sig(\gam\gam\to\ko\kob, J^P=0^+)$ we find
$(\epm\to\ks\ks\ {\rm or}\ \kl\kl)/(\epm\to\f\to\ks\kl)\sim
\pt{\rm few},-10,$

This gives us a unique opportunity to study: (a) \ks\ \BR's to high accuracy and (b) \ks\ rare decays: \ks\ semileptonic decays, \ks\to\pio\po, \ks\to\po$\nu\bar\nu$, in addition to \C\P\ and \C\P\T, the original mission of KLOE. Furthermore, KLOE can also study the neutral kaon system by ``interferometry''.

\FIG\kfifi
\section{KLOE's experimental program} 
Some topics of interest in KLOE's future are:
(1) measure kaon \C\P\ violating parameters, also by interference; (2) measure $V_{us}$, (3) find \ks\to\pio\po; (4) study \ks\to$\pi\ell\nu$; (5) verify $\Delta S=\Delta Q$; (6) keep an eye on \T\C\P, and (7) hopefully peek beyond the SM.
\subsection{Kaon \C\P\ violating parameters by kaon interferometry}
\setbox2=\vbox to.9cm{\vfil\hsize=15truecm \noindent\strut
\hbox{\rlap{\kern2.1truecm\rlap{$\bullet$}\kern
4truecm\rlap{$\bullet$}\kern 7truecm\rlap{$\bullet$}}\rlap{\raise
1.1truemm\hbox{\kern2.17truecm\vrule width11truecm height.7truept
depth.7truept}}\rlap{\raise 3truemm\hbox{\kern
4truecm\rlap{$t_1$}\kern5.5truecm\rlap{$t_2$}}}\rlap{\lower
3truemm\hbox{\kern3truecm\rlap{\ks, \kl}\kern5.5truecm\rlap{\kl,
\ks}}}\rlap{\kern1.2truecm\rlap{$f_1$}\kern 12.5truecm$
f_2$}\rlap{\raise 3.5truemm\hbox{\kern 6truecm$\f$}}}
\strut\vskip.3truecm\vfil }
\noindent
Interference allows the measurements of magnitude and phases of the kaon decay amplitudes without identifying long and short-lived kaons. Consider
the configuration illustrated
$$\box2$$
\allcap \kfifi;The decay \f\to\K\K\to$f_1,f_2$.;\vglue-2mm\noindent
The decay intensity is given by:
$$\eqalign{&I(f_1, f_2, t_1, t_2)=|
\langle\,f_1\sta{\ks}|^2|\langle\,f_2\sta{\ks}|^2e^{-\Gam_S\,t/2}\x\cr
&\quad[|\eta_1|^2e^{\Gam_S\Delta t/2}+|\eta_2|^2e^{-\Gam_S\Delta
t/2}-2|\eta_1||\eta_2|\cos(\Delta m\,
t+\phi_1-\phi_2)]\cr}.$$
Integrating over $t_1$ and $t_2$, for $\Delta t=t_1-t_2$ constant we find:
$$\eqalign{&I(f_1,\ f_2;\ \Delta t)=
{1\over2\Gamma}|\langle f_1\sta{\ks}\langle f_2\sta{\ks}|^2\x
[|\eta_1|^2e^{-\Gamma_L\Delta t}+\cr
&|\eta_2|^2e^{-\Gamma_S\Delta t}-
2|\eta_1||\eta_2|e^{-\Gamma\Delta t/2} \cos(\Delta m\Delta
t+\f_1-\f_2)]\cr }$$
We can thus measure $\Delta M$, $\Gamma$, $\eta_i$ phases.
($\eta_i={A(\kl\to i)/A(\ks\to i)}, \ \arg(\eta)=\phi$).
\subsection{Interference examples}
\FIG\inter
\cl{\figbox picpion;5;\hglue2cm\figbox ellelln;5;}
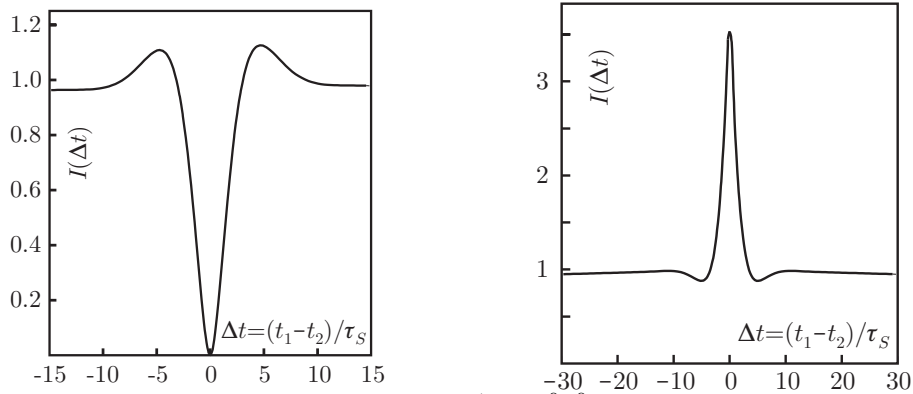
\allcap \inter; Interference patterns. Left
$f_{1,2}$=\pic,\pio, sensitive to \rep\ and $\Im(\eps'/\eps)$. 
Right $f_{1,2}=\ell^-,\ell^+$ yielding $\Re$ and $\Im$
of $A_{\ell^\pm}$;
\subsection{$\Gamma(\ks\to\pic)/\Gamma(\ks\to\pio)$ , KLOE 2001}
After all these years, chiral ``perturbationists'' are still trying to reach some accuracy in postdicting the $\Delta I=1/2$ rule. Its understanding would allow a more reliable calculation of \rep. Experimentally KLOE will measure the \BR's for \ks\ decays (and \kl) into various channels with high accuracy.
The ratio $R=\Gamma(\ks\to\pic)/\Gamma(\ks\to\pio)$ is not well known, only to a couple of \%, even though the double ratio $(\kl\to+-\kl\to00)/(\ks\to+-/\ks\to00)$ is known to \ab0.1\% . 

KLOE would like to reach 0.1\% on the former. Corrections are background sensitive. Luckily,  \kl\ interacting in the KLOE calorimeter gives an ideal \ks\ tag, almost independent of \ks\ decay mode. For our first measurement
of this ratio we obtain: ${R=2.239\pm0.003({\rm stat.})\pm0.015{(\rm syst.)}}$~\cite{kratioref}. Not only is this determination far more accurate than Particle Data Group' compilation, it is very important to note that KLOE includes all \ks\to\pic\gam, whereas the other previous experiments include unknown fractions, making comparison with theoretical calculations ambiguous. The above measurement is based on only 16 pb\up{-1}. KLOE in 2002 has about 30 times that amount of data to improve the accuracy of the measurement.

\FIG\ksemil
\FIG\kloeeep
\subsection{$\ks\to\pi e\nu$,  KLOE 2001}
\figboxc ksl3sigbw;6; 
\allcap \ksemil; The signal for \ks\to$e^\pm\pi^\mp\nu$ in KLOE.;\vglue.1mm\noindent
In the KLOE measurement of this process in 2001, we use only non spiraling tracks, time-of-flight, TOF, for electron identification and compare the missing energy, $E_{miss}$, with the missing momentum, $|p_{miss}|$, thus achieve an almost completely rejection of \pic\ backgrounds, as seen fig. \ksemil. We find: \BR=\pt(6.91\pm0.37),-4, ~\cite{klepref}. By end of 2002, KLOE will improve this measurement with 30 times more statistics, and will have the first measurement of ${\cal A}^S_\ell$. 
\def\tstrut{\vrule height 2.6ex depth .7ex}
\section{Expectations from the coming experiments}
\subsection{\rep\ via the double ratio method}
A review of the status of direct \C\P\ violation experiments has been given by B. Peyaud at this Rencontre, all references for NA48 and KTEV are in his article in this proceedings. Fig. \kloeeep\ shows the luminosity that KLOE needs to reach various accuracies on \rep\ by measuring the double ratio. KLOE's measurement will be performed at a \ff, with totally different systematics from the fixed target experiments. Furthermore, it will be charming, if nothing else, to see the interference patterns done with high accuracy.\\
\cl{\parbox{\textwidth}{\figboxc repinkloebw;11;\\
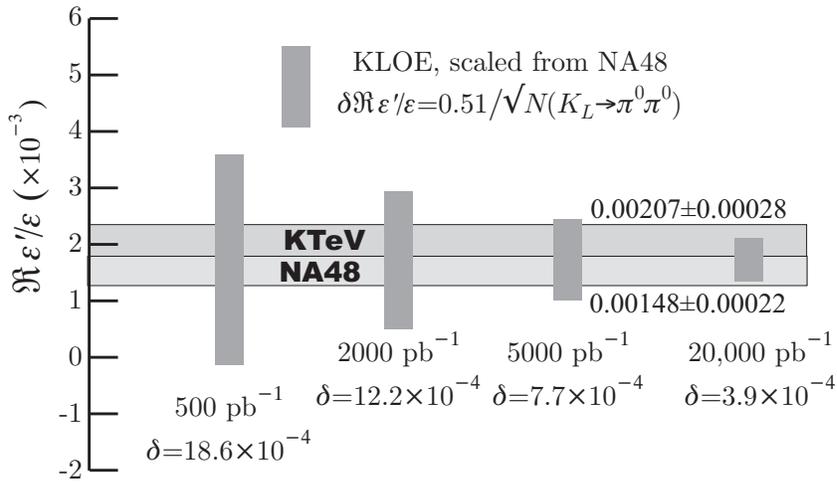
\allcap \kloeeep; KLOE expectations vs NA48 and KTeV.;\vglue-.5cm}}
\subsection{\kpm Decays}
Details and expectations of NA48/2 are given in Gianmaria Collazuol's paper. Table 2, is a summary of results expected from  NA48/2 and KLOE (Acc. stands for acceptance).\vglue2mm
\cl{Table 2. \kpm\ decays.}\vglue2mm
\begintable
 Mode | BR | Acc.       | Events | Acc. | Events | Note \crneg{1pt}
      |    |          &\hglue-1cm KLOE  |  & \hglue-1cm NA48/2    |   \crthick
\kpm\to\pic$\pi^\pm$ | 0.056 | 0.03 | \pt1.26,8, | |  \pt2,9,  | (1) \cr
\kpm\to\pio$\pi^\pm$ | 0.017 | 0.09 | \pt1.22,8, | | \pt1.2,8, | (2) \cr
\kpm\to$\pi^\pm$\po\gam | \pt2.8,-4, | 0.1 | \pt2,6, | 0.1 | \pt1,6, | (3) \cr
\kpm\to$\pi^\pm$\gam\gam\ | \pt5,-7, | 0.15 | 3750  |     |     | \endtable
\vglue2mm\cl{\parbox{12cm}{
(1). KLOE: $\delta A_g$=\pt4,-4,, $\delta A_\Gamma$=10\up{-4}. NA48/2: $\delta A_g$=\pt2,-4,.\\
(2). KLOE: $\delta A_g$=\pt2,-4,, $\delta A_\Gamma$=10\up{-4}. NA48/2: $\delta A_g$=\pt3.5,-4,.\\
(3). KLOE: $\delta A_\Gamma$=10\up{-3}. NA48/2 $\delta A_\Gamma$=10\up{-2}.}}

\subsection{\ks\ and \kl\ decays}
The experimental setups and expectations of NA48/1 are described in detail by Richard Batley in these proceedings. In table 3, I only summarize their anticipated results in comparison to those expected from KLOE (Acc. stands for acceptance).
\vglue2mm
\cl{Table 3. \ks\ and \kl\ decays.}\vglue2mm
 \begintable
 Mode | BR | Acc.       | Events | Acc. | Events | Note \crneg{1pt}
      |    |          &\hglue-1cm KLOE  |  & \hglue-1cm NA48/1    |   \crthick
 \ks\to\pic       | 0.67   | 0.15   | \pt5,9, |      |        |          \cr
 \ks\to\pio       | 0.31       | 0.15 | \pt2.5,9, |       |        |       \cr
 \ks\to$\pi e\nu$ | \pt7.4,-4, | 0.05 | \pt2,6,  |      |        |     \cr
 \ks\to\po\epm\  | \pt5.2,-9,      | 0.05 | 13   | 0.05 | 7  | Ind \nocp\ \cr
 \ks\to3\po\     | \pt2,-9,        | 0.17 | 16   | 0.05 | 4  | $\eta_{000}$ \cr
 \ks\to\po\gam\gam | \pt4,-8,    | 0.15 | 300 |  0.1   | 114   |    \cr
 \ks\to\pic\gam\ | \pt1.8,-3,  | 0.15 | \pt1.35,7, |    |    |     \cr
 \ks\to\pic\po | \pt3.2,-7, | 0.17 | 2500 |             |     |  \cr
  \kl\to\pic\po | 0.12 | 0.16 | \pt1,9, |      |        |          \cr
 \kl\to\pic    | 0.002 | 0.11 | \pt1.1,7, |       |        |          \cr
 \kl\to\pio    | 0.001 | 0.1  | \pt4,6, |       |        |          \cr
 \kl\to\pic\gam\ | \pt4.6,-5, | 0.16 | \pt3.7,5,  |      |        |   \endtable

\vglue1mm
\section{Conclusions}

NA48/1 and NA48/2 will (presumably) have results by 2004. KLOE will begin taking data in 2004. Thereafter one needs more intense kaon sources to attack the golden processes \kl\to\po$\nu\bar\nu$, \to\po\epm. After all these experiments are done there might still be some missing points: (a) will the origin of \nocp\ still be unknown? (b) \rep$\neq$0 rules out the superweak theory, but will the connection between CKM and \rep\ still evades us?
We have studied \nocp\ for 39 years, we still have quite a few more to go.

\section{Acknowledgements}

I want to thank J. Tran Thanh Van and Lydia Iconomidou-Fayard for organizing this most enjoyable and lively ``XIVth Rencontres DE BLOIS : {\it Matter-Antimatter Asymmetry}''
where so many of our colleagues could present and exchange ideas in a most convivial atmosphere.

\end{document}
